\documentclass{aa}
\usepackage{graphicx}
\usepackage{txfonts}
\begin{document}
   \title{Potential of colors for determining age and metallicity of stellar populations}


   \author{Zhongmu Li
          \inst{1, 2}
          \and
          Zhanwen Han\inst{1}
          \and
          Fenghui Zhang\inst{1}
          }

   \offprints{Zhongmu Li}

   \institute{National Astronomical Observatories/Yunnan Observatory, the Chinese Academy of Sciences, Kunming, 650011,
   China\\
              \email{zhongmu.li@gmail.com}
         \and
            Graduate School of the Chinese Academy of Sciences\\
             }
   \titlerunning {Colors for studying stellar populations}
   \date{Received September 04, 2006; accepted December 04, 2006}


  \abstract
   {Colors are usually not used for constraining stellar
   populations because they are thought to have the well-known age-metallicity degeneracy, but
   some recent works show that colors can also be used. A simple stellar
   population synthesis model is widely used,
   but there is no analysis for its colors.}
   {We try to find colors that can potentially be used to determine the age and
   metallicity of stellar populations by the standard model.}
   {Principal component analysis and relative sensitive
    parameter techniques are used in this work.}
   {$U-K$, $U-H$, $U-J$, $B-K$, $B-H$, $U-I$, $B-J$, and $V-K$ are found to be more important
    for studying populations than others. Pairs of colors such as $B-K$ and $B-V$
    are found to be able to disentangle the stellar age-metallicity degeneracy
    via the high-resolution model, while pairs such as $U-K$ and $R-I$ may be used instead when the
    low-resolution model is used. Furthermore, the $u-g$ and $r-i$ colors of the low-resolution model
    seem to have the same potential, but there are no such colors for the high-resolution one.}
   {Some colors have been shown to have the potential to determine the age and metallicity of stellar populations, but
   relative metallicity and age sensitivities of
   colors in different stellar population synthesis models are usually different. In addition, minor star formations
   will make star systems look younger and more metal rich than their dominating populations.}


   \maketitle
%

\section{Introduction}

    Determination of ages and metallicities of various stellar
    populations has long been a very important subject in astronomy and
    astrophysics because it can help us to understand not only
    galaxies, but also cosmology.
    Many such works are done via the popular technique called stellar population synthesis.
    Many simple stellar population (SSP) synthesis models, e.g.,
    Vazdekis (\cite{vazdekis}), Worthey (\cite{worthey}), Bruzual \&
    Charlot (\cite{bruzual}, hereafter BC03), and Zhang et al. (\cite{zhang}), are used widely and have shown us
    many new results.

    To study properties of stellar populations, most people choose to use
    spectra-like methods, e.g., absorption line indices, to determine
    age and metallicity as they have the ability to break
    the well-known age-metallicity degeneracy. But, in fact, we all hope to
    use colors for such work because they are easier
    to get than spectra and are independent of the distances of objects.
    A lot of work has been done in this way and recent such
    works are given by, e.g., Dorman et al. (\cite{dorman}) and James et al.
    (\cite{james}). The uncertainties of using
    synthetic integrated colors of SSPs to determine stellar ages are
    investigated by Yi (\cite{yi}).

    However, different stellar population synthesis models and colors are usually
    used by different authors and it seems difficult to do future work using the same
    method because other people do not have the same models.
    Therefore, it is valuable to study some widely used
    stellar population synthesis models. In this paper, we
    intend to analyze the BC03 standard model,
    with the aim of finding some colors that have the potential to
    break the degeneracy between stellar age and metallicity.

    The organization of the
    paper is as follows. In Sect. 2, we briefly introduce the BC03
    model and techniques used in this paper. In Sect. 3,
    we analyze $UBVRIJHK$ and $ugriz$ colors of the BC03 standard model.
    Finally, we give our conclusions in Sect. 4.


\section{BC03 model and techniques}

    The stellar population synthesis model of BC03 is a widely used model in stellar
    population synthesis. Its standard model takes the Chabrier (\cite {chabrier}) initial
    mass function (IMF) and uses Padova 1994 library tracks to calculate
    integrated colors. The model supplies us with both $UBVRIJHK$-kind
    and $ugriz$ colors that are compared to observations from the Sloan Digital
    Sky Survey (SDSS) and galactic clusters. Thus it is convenient to use
    the model for estimating stellar populations via colors.

    To search for colors, which are important to determine
    ages and metallicities of populations, we utilize two methods, i.e., principal component
    analysis (PCA) and relative metallicity sensitivity (RMS) techniques. PCA is a
    powerful technique for unveiling the correlation between
    variables in a data set and for determining the intrinsic
    dimensionality of a parameter space (Connolly \& Szalay \cite{connolly}). It
    can help us to reduce the dimension of variables and give the
    important variables in the initial data set. Therefore it can be
    used to find indices that have the potential to reveal
    properties of stellar populations (see Kong \&
    Cheng (\cite{kong}) for more detail).
    The algorithm works by building more compact and optimal linear
    combinations of data with respect to the mean square error
    criterion and calculating a new base with the minimum set of
    orthogonal axes, which are called principal components (PCs), and can describe the variance
    of the data on the whole. The most important PC is called the first
    principal component (PC1) and can usually express the main information of the original
    data. Thus the importance of each input variable for
    expressing the main information of the source data (now represented by PC1) can be estimated
    via the weights assigned to them, i.e., their correlations with
    PC1.
    In this work, we intend to find colors important to stellar age or metallicity via
    PCA. The results can, possibly, guide us to choose important colors to determine populations by
    multi-color methods and may also show us some colors that are
    markedly important to stellar age or metallicity, as with the spectral indices of Kong \& Cheng (\cite{kong}).

    As we will show, PCA of colors fails to show colors that are markedly important to stellar
    age or metallicity. We investigate the relative age and
    metallicity sensitivities of colors via the RMS technique,
    which can quantify how a color is sensitive to age and
    metallicity of SSPs. One can refer to Worthey (\cite{worthey}), in which RMSs for some colors and absorption line
    indices are quantified.


\section{PCA results and metallicity sensitivities of colors}

\subsection{Main results}

    To find colors important to stellar age, we analyze the
    colors of SSPs with the same metallicity and check their correlations to PC1,
    as PC1 expressed more than 94$\%$ of the information from the original data.
    Because the differences in colors result purely from
    stellar age and are expressed mainly by PC1, colors that correlate strongly to PC1 will be
    important to stellar age. A similar method is also used to
    look for colors important to stellar metallicity.
    Because $UBVRIJHK$ and $ugriz$ colors can usually be derived from many
    sources (e.g., SDSS and Two-Micron All-Sky Survey (2MASS)) and seem to be more available
    than others, we analyze them in this work. The main results
    for $UBVRIJHK$ colors of high-resolution SSPs are shown in Tables 1 and 2. In each
    table, the
    contribution of PC1 to the total variance and the weight assigned to each input color are shown.
    As the results for colors of
    low-resolution SSPs are very similar to those of the high-resolution ones, we do not list them here.
    We do not show those for $ugriz$ colors either, as their PCA results are similar.

\begin{table}[]
\caption[]{PCA results of $UBVRIJHK$ colors of SSPs in the
high-resolution BC03 model. Each row except the first shows the
results for colors of SSPs with the same metallicity. ``Percent''
and ``Correlations'' are the contribution of PC1 to the total
variance and correlations of colors to PC1.} \label{Tab:1}
\begin{center}\begin{tabular}{lcccc}
\hline \hline\noalign{\smallskip}
Z          &0.0001    &0.0040 &0.0200    &0.0500  \\    
\hline
           &PC1       &PC1       &PC1       &PC1       \\
Percent    &99.72     &98.33     &97.54     &94.40     \\
\hline
  &&Correlations&&  \\
\hline                        
        U-K&-0.376    &-0.378    &-0.379    &-0.379      \\
        U-H&-0.363    &-0.364    &-0.365    &-0.364      \\
        U-J&-0.320    &-0.321    &-0.323    &-0.326      \\
        B-K&-0.287    &-0.287    &-0.285    &-0.284      \\
        B-H&-0.274    &-0.273    &-0.271    &-0.269      \\
        U-I&-0.244    &-0.248    &-0.243    &-0.244      \\
        B-J&-0.231    &-0.230    &-0.229    &-0.230      \\
        V-K&-0.217    &-0.214    &-0.214    &-0.215      \\
        U-R&-0.202    &-0.207    &-0.204    &-0.205      \\
        V-H&-0.204    &-0.200    &-0.200    &-0.199      \\
        R-K&-0.174    &-0.171    &-0.175    &-0.174      \\
        U-V&-0.159    &-0.164    &-0.165    &-0.164      \\
        R-H&-0.161    &-0.157    &-0.161    &-0.159      \\
        V-J&-0.161    &-0.157    &-0.158    &-0.161      \\
        B-I&-0.155    &-0.157    &-0.148    &-0.148      \\
        I-K&-0.132    &-0.130    &-0.136    &-0.135      \\
        I-H&-0.119    &-0.116    &-0.122    &-0.120      \\
        R-J&-0.118    &-0.114    &-0.119    &-0.121      \\
        B-R&-0.113    &-0.116    &-0.110    &-0.109      \\
        U-B&-0.089    &-0.091    &-0.095    &-0.095      \\
        V-I&-0.086    &-0.084    &-0.078    &-0.079      \\
        I-J&-0.076    &-0.073    &-0.081    &-0.082      \\
        B-V&-0.070    &-0.073    &-0.071    &-0.069      \\
        J-K&-0.056    &-0.057    &-0.056    &-0.053      \\
        J-H&-0.043    &-0.043    &-0.042    &-0.038      \\
        V-R&-0.043    &-0.043    &-0.039    &-0.040      \\
        R-I&-0.042    &-0.041    &-0.038    &-0.039      \\
        H-K&-0.013    &-0.014    &-0.014    &-0.015      \\

\noalign{\smallskip}\hline
\end{tabular}\end{center}
\end{table}

\begin{table}[]
\caption[]{Similar to Table 1, but each row represents PCA result
for colors of SSPs with the same age.} \label{Tab:2}
\begin{center}\begin{tabular}{lcccc}
\hline \hline\noalign{\smallskip}
t/Gyr      &2    &8   &14   &20       \\
\hline
           &PC1       &PC1       &PC1       &PC1       \\
Percent    &99.73   &   99.86&   99.84&   99.81   \\

\hline
  &&Correlations&&   \\
\hline
        U-K&-0.370    &-0.384    &-0.387    &-0.389    \\
        U-H&-0.343    &-0.361    &-0.366    &-0.370    \\
        U-J&-0.307    &-0.326    &-0.334    &-0.341    \\
        B-K&-0.305    &-0.283    &-0.277    &-0.269    \\
        B-H&-0.278    &-0.261    &-0.256    &-0.251    \\
        U-I&-0.197    &-0.226    &-0.240    &-0.248    \\
        B-J&-0.242    &-0.226    &-0.224    &-0.221    \\
        V-K&-0.237    &-0.222    &-0.213    &-0.206    \\
        U-R&-0.165    &-0.193    &-0.206    &-0.214    \\
        V-H&-0.210    &-0.200    &-0.192    &-0.187    \\
        R-K&-0.206    &-0.191    &-0.181    &-0.175    \\
        U-V&-0.133    &-0.162    &-0.174    &-0.183    \\
        R-H&-0.178    &-0.168    &-0.160    &-0.156    \\
        V-J&-0.174    &-0.165    &-0.160    &-0.157    \\
        I-K&-0.174    &-0.158    &-0.147    &-0.141    \\
        R-J&-0.143    &-0.134    &-0.128    &-0.126    \\
        I-H&-0.147    &-0.136    &-0.126    &-0.123    \\
        B-I&-0.132    &-0.125    &-0.130    &-0.128    \\
        U-B&-0.065    &-0.100    &-0.110    &-0.120    \\
        I-J&-0.111    &-0.101    &-0.094    &-0.093    \\
        B-R&-0.100    &-0.093    &-0.096    &-0.095    \\
        V-I&-0.063    &-0.064    &-0.066    &-0.064    \\
        B-V&-0.068    &-0.062    &-0.064    &-0.064    \\
        J-K&-0.063    &-0.057    &-0.053    &-0.048    \\
        R-I&-0.032    &-0.033    &-0.034    &-0.033    \\
        J-H&-0.036    &-0.035    &-0.032    &-0.030    \\
        V-R&-0.032    &-0.031    &-0.032    &-0.031    \\
        H-K&-0.027    &-0.022    &-0.021    &-0.019    \\

\noalign{\smallskip}\hline
\end{tabular}\end{center}
\end{table}

    Tables 1 and 2 show the contribution of each PC1 for expressing the original information and correlations of colors to PC1,
    with a mainly descending order. Colors with big correlations are more important to age and metallicity than those with
    small correlations. In detail, $U-K$, $U-H$, $U-J$, $B-K$, $B-H$,
    $U-I$, $B-J$, and $V-K$ colors are shown to be more important than others to determine the ages and metallicities of
    stellar populations. In addition, most colors
    that are important to stellar metallicity
    are important to stellar age. Thus it is
    difficult to determine stellar age or metallicity via only a color.
    In other words, it is more appropriate to determine stellar age and metallicity of a
    population via a few colors. This result is in agreement with that of Chang et al. (\cite{chang}) that
    optical colors of elliptical galaxies are sensitive to a combination of age, metallicity, and
    $\alpha$-enhancement.

    As we want to find pairs of colors that are able to disentangle the age-metallicity degeneracy,
    we investigate the relative age and metallicity sensitivities
    of colors using a method similar to Worthey (\cite{worthey}).
    A RMS is defined for each color to estimate how it is sensitive to age and metallicity.
    In detail, RMS is the ratio of
    percentage change of age to that of metallicity
    when they lead to the same change in a color, respectively. According to
    the definition, colors with large RMSs ($>$1.0) are more sensitive to metallicity
    and those with small RMSs ($<$1.0) are more sensitive to stellar age.
    In this work, we do not fix the zero point (see Worthey (\cite{worthey}) for comparison).
    The main results are shown in Table 3, in which the average RMSs
    of $UBVRIJHK$ and $ugriz$ colors are listed. The first and third
    columns show the results for colors of the high-resolution BC03 model and the second and fourth for the
    low-resolution model.

\begin{table}[]
\caption[]{Relative metallicity sensitivities of colors. The left
two columns show results for $UBVRIHJK$ and the right two for
$ugriz$ colors. 'High' and 'Low' represent high and low resolution
models of BC03, respectively.} \label{Tab:1}
\begin{center}\begin{tabular}{lc|lc||lc|lc}
\hline\hline\noalign{\smallskip}
&High&&Low&&High&&Low\\
\hline
      B-K &2.7029&U-K &2.7077&u-z &1.7542&u-g &1.7598\\
      R-K &2.2928&I-H &2.3703&u-i &1.5477&i-z &1.4833\\
      I-H &2.2870&R-J &2.0011&r-z &1.5225&u-i &1.4609\\
      I-J &2.0531&R-H &1.9113&i-z &1.4683&u-z &1.4014\\
      U-K &2.0250&R-K &1.8018&g-i &1.4370&u-r &1.3825\\
      V-K &1.9098&U-R &1.7760&g-z &1.3907&r-z &1.3193\\
      V-H &1.7760&I-J &1.7533&g-r &1.2722&g-z &1.1709\\
      I-K &1.7017&U-H &1.7165&u-r &1.2659&g-r &1.0959\\
      U-H &1.6686&V-H &1.7116&r-i &1.1425&g-i &1.0437\\
      V-J &1.6544&I-K &1.6924&u-g &1.0605&r-i &0.7936\\
      R-J &1.6201&V-K &1.6696&    &           &      \\
      U-J &1.5372&B-J &1.6425&    &           &      \\
      R-H &1.4821&U-J &1.6153&    &           &      \\
      U-I &1.3298&B-K &1.5934&    &           &      \\
      B-I &1.2154&U-I &1.5452&    &           &      \\
      B-J &1.1785&B-H &1.5184&    &           &      \\
      B-H &1.1669&U-V &1.3781&    &           &      \\
      U-B &0.9862&B-I &1.3367&    &           &      \\
      B-R &0.9483&V-J &1.2083&    &           &      \\
      V-R &0.7973&B-V &1.0092&    &           &      \\
      U-V &0.7362&B-R &0.9572&    &           &      \\
      J-H &0.7339&J-H &0.9439&    &           &      \\
      V-I &0.7132&V-I &0.7675&    &           &      \\
      J-K &0.6677&U-B &0.7254&    &           &      \\
      H-K &0.5506&V-R &0.7137&    &           &      \\
      R-I &0.4522&H-K &0.5506&    &           &      \\
      U-R &0.3974&J-K &0.5290&    &           &      \\
      B-V &0.3695&R-I &0.4417&    &           &      \\
\noalign{\smallskip}\hline
\end{tabular}\end{center}
\end{table}

    The results suggest that RMSs of colors in high- and
    low-resolution models are different, as we see in Table 3. In the
    high-resolution model $B-K$ and $B-V$, but for the low-resolution model it is $U-K$ and $R-I$ that are
    sensitive to stellar metallicity and age, respectively.
    Moreover, $u-g$ and $r-i$ colors of the low-resolution model are found to be
    sensitive to stellar metallicity and age, respectively, but they show
    different sensitivities in the high-resolution model.
    As a whole, our result agrees well with previous works that
    the age-metallicity degeneracy can be disentangled by using optical and infrared colors together.
    One can refer to previous work of
    James et al. (\cite{james}), in which $J-K$ and $B-K$
    colors are used as the metallicity and age sensitive colors and the work of Peletier
    et al. (\cite{peletier}), in which $U-V$, $B-V$, and $V-K$ are used to
    constrain the stellar populations of 12 galaxies.

\subsection{How colors can disentangle the age-metallicity degeneracy}

    We plot some pairs of colors suggested by the above results to
    check their abilities for disentangling the stellar age-metallicity degeneracy and for
    constraining stellar populations, which can be
    seen in Figs. 1, 2, and 3. The three figures plot pairs of $UBVRIJHK$ colors suggested
    for the high- and low-resolution models and $ugriz$ colors for the low-resolution model,
    respectively. In Fig. 1, $B-V$ and $B-K$, which are most likely to break the age-metallicity
    degeneracy via colors of the high-resolution BC03 model, are plotted.
    Constant age (isochrones) and constant metallicity
    are represented differently. It seems that the two colors have
    some potential to distinguish stellar populations. In Figs. 2 and 3, the similar relations for colors of the low-resolution BC03 model,
    i.e., $R-I$ versus $U-K$ and $r-i$ versus $u-g$ are plotted. We noticed that
    the AB-system colors suggested for the high-resolution model seem unable to separate the effects of
    age and metallicity and are not shown here.

   \begin{figure}
   \centering
   \includegraphics[angle=-90,width=88mm]{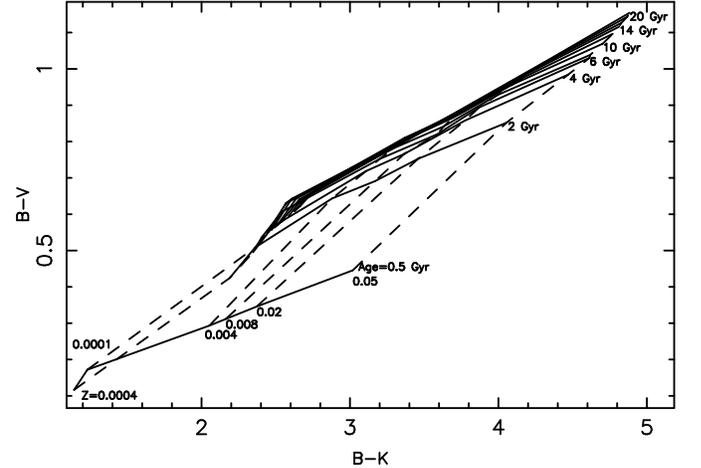}
      \caption{$B-V$ vs. $B-K$ plane predicted by the high-resolution BC03 model.
      Solid and dashed lines represent constant age and constant
      metallicity,
      respectively.
              }
         \label{FigVibStab}
   \end{figure}
%

   \begin{figure}
   \centering
   \includegraphics[angle=-90,width=88mm]{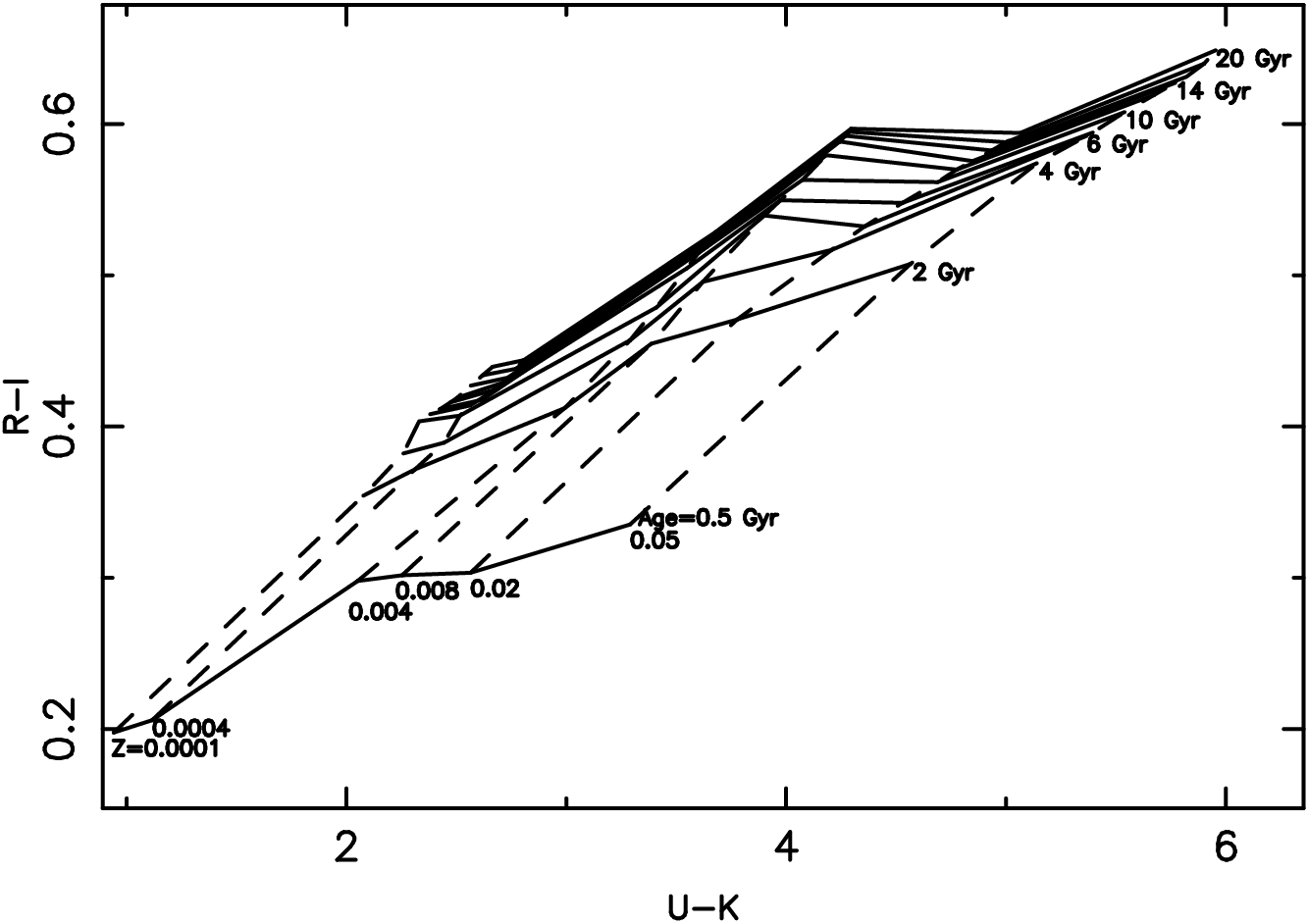}
      \caption{$R-I$ vs. $U-K$ plane predicted by the low-resolution BC03
      model. Lines have the same meanings as in Fig. 1.
              }
         \label{FigVibStab}
   \end{figure}
%

   \begin{figure}
   \centering
   \includegraphics[angle=-90,width=88mm]{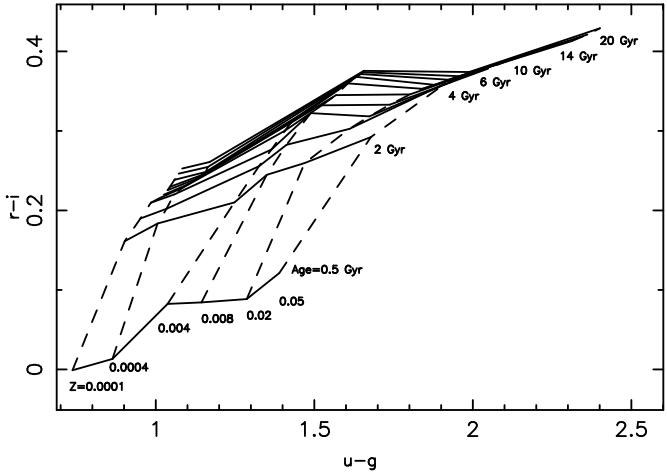}
      \caption{$r-i$ vs. $u-g$ plane predicted by the low-resolution BC03
      model. Lines have the same meanings as in Fig. 1.
              }
         \label{FigVibStab}
   \end{figure}
%

    To compare with observation, we plot some observed galaxies in the $B-V$ versus $B-K$ plane
    and $r-i$ versus $u-g$ plane, which can be seen in Figs. 4 and 5.
    The first sample is selected from Michard (\cite{michard}), consisting of 53 elliptical galaxies that have
    correct $UBVRIJHK$ photometry supplied and the second one from the
    Sloan Digital Sky Survery Data Release Five (SDSS-DR5), consisting of 12901 galaxies with
    concentration index C $\geq$ 2.8, absolute $r$-band Petrosian magnitude $M_{r}$ $\leq$ -22,
    redshift 0.1 $\leq$ $\emph {z}$ $<$ 0.25, $r$-band Petrosian magnitude 14.5 $<$ $r$ $<$ 17.77, and
    with the best quality in ``Photoz''. The $k$-correction, galactic
    extinction, and differences between SDSS $ugriz$ magnitudes and the AB-system are taken into
    account using the data supplied by SDSS, in a cosmology with $\Omega$ = 0.3, $\Lambda$ =0.7, and $H_{0}$ =
    70 km s$^{-1} $Mpc$^{-1}$. We see that some galaxies are out of the
    grids, which perhaps results from the limit of the theoretical model and ongoing star formations.
    The detailed parameters of galaxies in the first sample are shown in Table 4,
    in which best-fitted metallicities, ages, and their
    associated 1 $\sigma$ uncertainties of galaxies are listed. The results show that stellar ages and metallicities of these galaxies can
    be disentangled using the BC03 model, with average relative uncertainties of 0.26 and 0.13 in age and metallicity, respectively.
    However, most of these galaxies are shown to be very young and richer than solar metallicity,
    which possibly results from minor star formations in
    them. In fact, recent star formations can change colors of a
    population significantly. Here we test the changes in Fig. 6 by
    mixing a young stellar population (0.5 Gyr) into an old population (15
    Gyr) with different percentages. Two stellar populations have the same metallicity. The numbers from 0 to 8 in the figure show that the young
    population contributes 0, 0.5, 1.0, 2.0, 3.0, 4.0, 10.0, 20.0, and 100 per cent
    in mass to the total system. As seen in Fig. 6, a young stellar
    population, even if it contributes to the whole system by only
    0.5 per cent, will change the colors of the star system significantly, which makes the system
    look more metal rich and younger. This result is similar to that of line-strength indices of such systems
    (see,
    e.g., Serra \& Trager \cite{serra}). Therefore, the presence of galaxies that are out of the grid may result from recent star
    formations, and the dominating stellar populations of our sample galaxies, which contribute the most mass to these systems, are
    possibly older and less metal rich than that shown in Table 4. Our work also shows that if the young stellar population is older than about 1 Gyr and takes only a
    small percentage, the metallicities measured from $B-K$ and $B-V$ are similar to their
    dominating populations, while the derived ages bias to their younger populations (see Fig. 7 in more detail).
    Thus the metallicities of our sample galaxies are credible and can be used for future studies.

\begin{table}[]
\caption[]{Stellar ages and metallicities of 53 elliptical galaxies
and the associated 1 $\sigma$ uncertainties, fitted by BC03 using
$B-K$ and $B-V$ colors.} \label{Tab:1}
\begin{center}\begin{tabular}{l|r|r|l|r}
\hline\hline\noalign{\smallskip}
Name       & Age      &Error    &Z            & Error\\
           & [Gyr]    &[Gyr]    &             &      \\
\hline
NGC0584      &3.126     &0.720     &0.038            &0.005\\
NGC0596      &4.416     &1.066     &0.029            &0.003\\
NGC0720      &7.762     &4.540     &0.022            &0.006\\
NGC0821      &5.370     &1.086     &0.035            &0.003\\
NGC1052      &3.758     &1.196     &0.039            &0.004\\
NGC1395      &3.802     &1.153     &0.043            &0.005\\
NGC1399      &5.070     &1.179     &0.042            &0.004\\
NGC1404      &5.821     &2.031     &0.037            &0.004\\
NGC1407      &3.126     &0.463     &0.044            &0.006\\
NGC1537      &2.786     &0.845     &0.044            &0.010\\
NGC1549      &3.055     &0.453     &0.046            &0.006\\
NGC1600      &3.388     &1.130     &0.048            &0.005\\
NGC1700      &2.786     &0.304     &0.044            &0.005\\
NGC2768      &3.350     &2.021     &0.031            &0.009\\
NGC2974      &4.315     &1.042     &0.037            &0.005\\
NGC2986      &3.311     &0.149     &$\geq$0.05       &     \\
NGC3115      &2.951     &0.360     &0.049            &0.005\\
NGC3193      &5.070     &1.025     &0.030            &0.003\\
NGC3250      &2.818     &0.237     &$\geq$0.05       &     \\
NGC3377      &4.074     &0.996     &0.023            &0.003\\
NGC3379      &3.055     &0.295     &$\geq$0.05       &     \\
NGC3557      &2.570     &0.628     &0.048            &0.008\\
NGC3608      &5.248     &1.283     &0.029            &0.003\\
NGC3610      &1.603     &0.216     &0.049            &0.004\\
NGC3962      &3.311     &1.877     &0.044            &0.007\\
NGC4125      &3.508     &2.116     &0.030            &0.010\\
NGC4261      &6.026     &2.010     &0.035            &0.004\\
NGC4278      &2.917     &0.281     &0.037            &0.004\\
NGC4365      &3.758     &1.196     &0.039            &0.004\\
NGC4374      &3.236     &1.079     &0.039            &0.005\\
NGC4406      &4.898     &1.052     &0.031            &0.004\\
NGC4472      &3.388     &0.119     &$\geq$0.05       &     \\
NGC4473      &2.754     &0.271     &0.039            &0.005\\
NGC4478      &2.851     &0.251     &0.030            &0.004\\
NGC4486      &3.090     &0.139     &$\geq$0.05       &     \\
NGC4494      &2.630     &0.339     &0.036            &0.004\\
NGC4552      &5.129     &0.967     &0.035            &0.003\\
NGC4564      &5.689     &2.347     &0.023            &0.005\\
NGC4589      &2.113     &0.487     &0.049            &0.004\\
NGC4621      &3.055     &0.453     &0.046            &0.006\\
NGC4636      &2.884     &1.333     &0.041            &0.009\\
NGC4649      &6.310     &2.008     &0.038            &0.003\\
NGC4660      &3.802     &1.153     &0.043            &0.005\\
NGC4696      &3.350     &1.959     &0.046            &0.008\\
NGC4697      &2.985     &1.585     &0.035            &0.006\\
NGC5322      &1.928     &0.286     &$\geq$0.05       &     \\
NGC5576      &2.188     &0.295     &0.040            &0.004\\
NGC5813      &4.365     &1.054     &0.039            &0.005\\
NGC5846      &5.070     &2.965     &0.037            &0.008\\
NGC5866      &2.213     &0.186     &$\geq$0.05       &     \\
NGC7144      &3.846     &1.052     &0.033            &0.004\\
NGC7507      &3.936     &1.019     &0.035            &0.004\\
 IC1459      &4.898     &1.225     &0.039            &0.004\\

\noalign{\smallskip}\hline
\end{tabular}\end{center}
\end{table}

   \begin{figure}
   \centering
   \includegraphics[angle=-90,width=88mm]{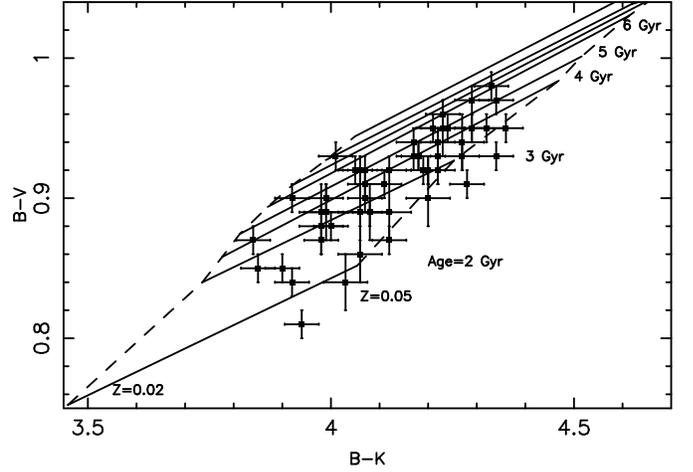}
      \caption{Colors of 53 elliptical galaxies overlaid onto the $B-V$ vs. $B-K$
      plane. Error bars show the uncertainties. Solid and dashed lines have the same meanings as in Fig. 1.
              }
         \label{FigVibStab}
   \end{figure}
%

   \begin{figure}
   \centering
   \includegraphics[angle=-90,width=88mm]{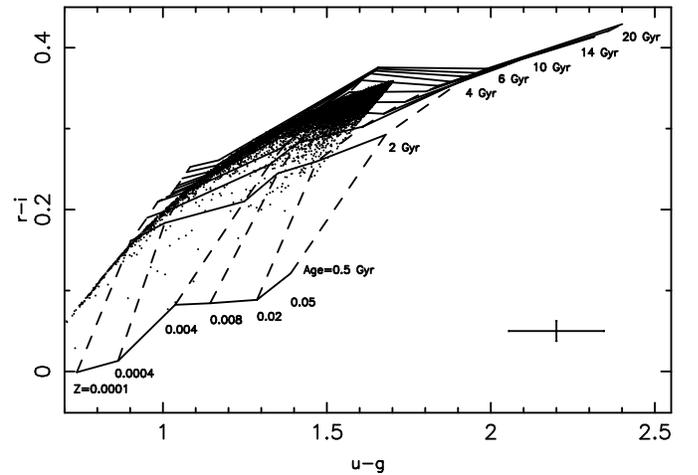}
      \caption{$r-i$ and $u-g$ colors of 12901 early-type galaxies overlaid onto the
      theoretical calibration. Error bars show the typical uncertainties. Solid and dashed lines have the same meanings as in Fig. 1.
              }
         \label{FigVibStab}
   \end{figure}
%


\section{Conclusions}

    We analyzed $UBVRIJHK$ and $ugriz$ colors of high- and
    low-resolution models of BC03 via PCA and RMS
    techniques. The PCA show that colors such as $U-K$, $U-H$, $U-J$, $B-K$, $B-H$,
    $U-I$, $B-J$, and $V-K$ colors are better than others for
    determining stellar populations via multi-color methods. The RMS results show that RMSs of colors in different SSP models are very different from
    each other, even for the same model with just two different resolutions.
    For the high-resolution model of BC03, $B-K$ and $B-V$ are found
    to be sensitive, respectively, to stellar metallicity and age, and are
    possibly suitable for constraining populations.
    $U-K$ and $R-I$ should be used instead for the
    low-resolution model. In addition, $u-g$ and $r-i$ colors of the
    low-resolution model are shown to
    have the potential to constrain the stellar metallicity and age of
    populations. The results also show that it is better to use
    optical and infrared colors together for determining
    populations. However, if there are ongoing star formations, we will get smaller ages and richer metallicities
    than their dominating populations.

   \begin{figure}
   \centering
   \includegraphics[angle=-90,width=88mm]{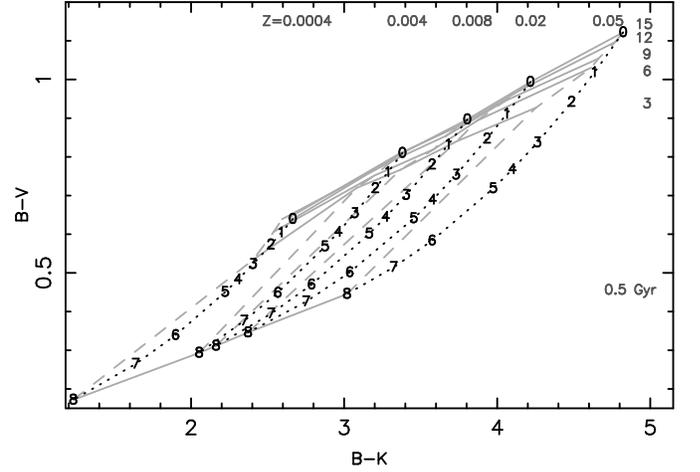}
      \caption{Changes of $B-K$ and $B-V$ colors when mixing a young stellar population into an old
      one. The two stellar populations have the same metallicity.
      Solid and dashed lines are same as in Fig. 1. Dotted lines represent constant metallicity.
      Numbers from 0 to 8 show colors of an old population (15 Gyr) mixed with a young population (0.5
      Gyr) with different percentages. From 0 to 8, the young population contributes 0, 0.5, 1.0, 2.0, 3.0, 4.0, 10.0,
      20.0, and 100 per cent in mass to the total system.
              }
         \label{FigVibStab}
   \end{figure}
%

   \begin{figure}
   \centering
   \includegraphics[angle=-90,width=88mm]{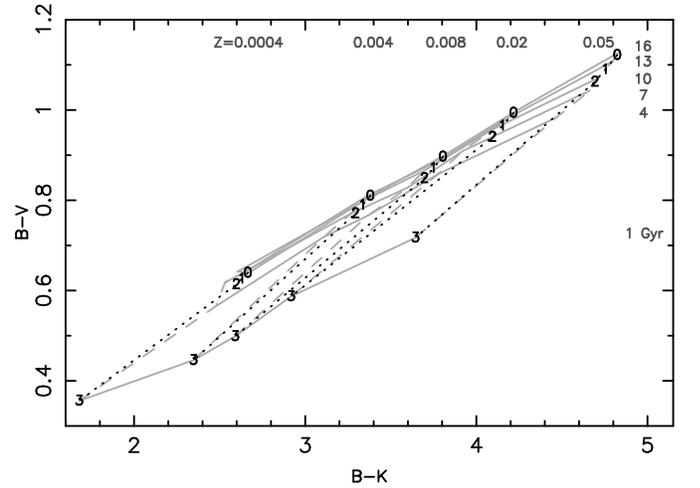}
      \caption{Similar to Fig. 6, but for a composite system which
      includes a 15 Gyr stellar population and a 1 Gyr stellar
      population. Numbers from 0 to 3 mean that the young population contributes 0, 0.5,
      1.0, and 100 per cent in mass to the total system.
              }
         \label{FigVibStab}
   \end{figure}
%

\begin{acknowledgements}
We gratefully acknowledge Dr.~ Richard Simon Pokorny for checking
the English and Prof.~Xu Kong for some useful discussions. We also
thank the anonymous referee for useful comments that helped clarify
some important points of the paper. This work is supported by the
Chinese National Science Foundation (Grant Nos. 10433030, 10521001,
and 10303006), the Chinese Academy of Sciences (No. KJX2-SW-T06),
and Yunnan Natural Science Foundation (Grant No. 2005A0035Q).
\end{acknowledgements}

\end{document}